\documentclass[a4paper]{iopart}

\usepackage[dvips]{graphicx}
\usepackage{times}
\usepackage{latexsym}
\usepackage{amssymb}
\usepackage{color}

\newcommand{\cf}{\textit{cf.}~}
\newcommand{\ie}{\textit{i.e.,}}
\newcommand{\eg}{\textit{e.g.,}}

\newcommand{\eq}{\begin{equation}}
\newcommand{\eeq}{\end{equation}}
\newcommand{\be}{\begin{equation}}
\newcommand{\ee}{\end{equation}}
\newcommand{\bea}{\begin{eqnarray}}

\newcommand{\eea}{\end{eqnarray}}

\newcommand{\vta}[1]{\vert \mathbf{a}_{#1}         \vert}
\newcommand{\vtaf}  {\vert \mathbf{a}_{\rm fin}     \vert}
\newcommand{\vtl}{\vert \ell \vert}

\newcommand{\boldell}{{\boldmath $\ell$}~}

\begin{document}

\title[Modelling the final state from binary black-hole
  coalescences]{Modelling the final state from binary black-hole
  coalescences}

\author{Luciano Rezzolla
}

\address{Max-Planck-Institut f\"ur Gravitationsphysik, Albert Einstein
Institut, %\\Am Muehlenberg 1, 14476
Golm, Germany}

\address{Department of Physics, Louisiana State University, Baton
  Rouge, LA 70803 USA}

\begin{abstract}
Over the last few years enormous progress has been made in the
numerical description of the inspiral and merger of binary black
holes. A particular effort has gone into the modelling of the physical
properties of the final black hole, namely its spin and recoil
velocity, as these quantities have direct impact in astrophysics,
cosmology and, of course, general relativity. As numerical-relativity
calculations still remain computationally very expensive and cannot be
used to investigate the complete space of possible parameters,
semi-analytic approaches have been developed and shown to reproduce
with very high precision the numerical results. I here collect and
review these efforts, pointing out the relative strengths and
weaknesses, and discuss which directions are more promising to
further improve them.
 \end{abstract}
\maketitle

%-------------------------------------------------------------%
\section{Introduction}
\label{sect0}
%-------------------------------------------------------------%

Despite the almost unnatural simplicity with which the problem can be
formulated (black holes are after all the simplest macroscopical
objects we know), the final evolution of a binary system of black
holes is an impressively complex problem to solve. At the same time,
this very simple process plays a fundamental role in astrophysics, in
cosmology, in gravitational-wave astronomy, and of course in general
relativity. Recent progress in numerical relativity initiated by the
works in refs.~\cite{Pretorius:2005gq, Campanelli:2005dd,
  Baker:2006yw}, have made it now possible to compute the different
stages of the evolution, starting from the inspiral at large
separations, for which post-Newtonian (PN) calculations provide an
accurate description, through the highly relativistic merger, and
finally to the ringdown.

As long as the two black holes are not extremal and have masses which
are not too different from each other (see
however~\cite{Gonzalez:2008} for simulations with a mass-ratio of
$1\!\!:\!\!10$), no major technical obstacle now prevents the solution
of this problem in full generality and with an overall error which 
can be brought down to less than $1\%$~ (see
ref.~\cite{Pollney:2007ss, Baker:2008, Campanelli:2008, Hinder:2008kv,
  Gonzalez:2008} for some recent examples). Yet, obtaining such a
solution still requires a formidable computational power sustained
over several days. Even for the simplest set of initial data, namely
those considering two black holes in quasi-circular orbits, the space
of parameters is too vast to be explored entirely through
numerical-relativity calculations. Furthermore, many studies of
astrophysical interest, such as many-body simulations of galaxy
mergers, or hierarchical models of black-hole formation, span a
statistically large space of parameters and are only remotely
interested in the evolution of the system during the last few tens of
orbits and much interested in determining the properties of the final
black hole when the system is still widely separated.

In order to accommodate these two distinct and contrasting needs,
namely that of sampling the largest possible space of parameters and
that of reducing the computational costs, a number of analytical or
semi-analytic approaches have been developed over the last couple of
years.  In most of these approaches the inspiral and merger is
considered as a process that takes, as input, two black holes of
initial masses $M_{1}$, $M_{2}$ and spin vectors {\boldmath
  $S$}$_{1}$, {\boldmath $S$}$_{2}$ and produces, as output, a third
black hole of mass $M_{\rm fin}$, spin {\boldmath $S$}$_{\rm fin}$ and
recoil velocity {\boldmath $v$}$_{\rm kick}$. Mathematically,
therefore, one is searching for a mapping between the initial
7-dimensional space of parameters (\ie~the one containing the six spin
components $S^j_{1,2}$ and the mass ratio $q \equiv M_2/M_1$) to two
distinct 3-dimensional spaces (\ie~the ones containing the three
components of the final spin and of the recoil velocity)\footnote{The
  space of final parameters could in principle be larger if the final
  mass of the black hole were to be modelled. Two attempts to do so
  have been recently proposed~\cite{Kesden:2008,Tichy:2008} (see also
  the discussion in Sec.~\ref{sect1}), but this quantity is not
  usually modelled.}. Clearly this is a degenerate mapping (two
different initial configurations can lead to the same final one) and
it would seem a formidable task to accomplish given the highly
nonlinear features of the few last orbits. Yet surprisingly, or
perhaps not so surprisingly given the underlying simplicity of the
problem, all of these studies have shown that the final spin vector
and the final recoil velocity vector, can be estimated to remarkably
good accuracy if the initial parameters are
known~\cite{Rezzolla-etal-2007, Buonanno:07b, BoyleKesdenNissanke:07,
  BoyleKesden:07, Marronetti:2007wz, Rezzolla-etal-2007b,
  Rezzolla-etal-2007c, Kesden:2008, Tichy:2008}.

This paper is dedicated to review the different approaches developed
so far to tackle this problem, pointing out the relative strengths and
weaknesses, and finally comparing them against the numerical data.
The manuscript is organized as follows: in Sect.~\ref{sect1} I review
the different methods that have been suggested to model the final
spin, distinguishing those that are purely analytic from those that
employ, with different amounts, the results of numerical
simulations. Sect.~\ref{sect2} is instead dedicated to the specific
approach developed at the AEI to obtain a simple and robust algebraic
expression for the final spin vector. In Sect.~\ref{sect2} I also
exploit the advantages of such analytic formula to explore those
configurations that may be astrophysically more interesting and then
validate the accuracy of the formula against all of the available data
and across alternative approaches. Finally, Sect.~\ref{sect3} is
dedicated to the modelling of the recoil velocity vector and a summary
is presented of the different contributions and of the debate still
present on one them. Section~\ref{sect4} will present the conclusions
and highlight the prospects of future work in the modelling of the
final state.

As a final but important remark I note that all of the considerations
made here apply to binary systems that inspiral from very large
separations and hence through quasi-circular orbits. Such
configurations are the ones more likely to occur astrophysically since
any residual eccentricity is lost quickly by the
gravitational-radiation reaction~\cite{Peters:1964}. Furthermore, at
least for nonspinning equal-mass black holes, recent work has shown
that the final spin does not depend on the value of the eccentricity
as long as the latter is not too large~\cite{Hinder:2007qu}.

%-------------------------------------------------------------%
\section{Modelling the final spin vector}
\label{sect1}
%-------------------------------------------------------------%

A number of analytical approaches have been developed over the years
to determine the final spin from a binary black hole
coalescence~\cite{Buonanno00a, Damour:2001tu, Buonanno:06cd,
  DamourNagar:07a, Gergely:08}. A particularly active line of research
is the one that has exploited the motion of test particles in black
hole spacetimes. A first attempt in this direction was made by Hughes
and Blandford~\cite{Hughes:2002ei}, who assumed that the energy and
angular momentum radiated in the merger and ringdown phase is much
smaller than the one radiated during the inspiral. In this way they
were able to express the conservation of energy and angular momentum
when the two black holes reach the innermost stable circular orbit
(ISCO) and hence compute the mass and spin of the final black
hole\footnote{I note that although numerical simulations do not show
  evidence for the existence of an ISCO, the concept of an effective
  ISCO can nevertheless be useful for the construction of
  gravitational-wave templates~\cite{Ajith:2007kx,Hanna2008}.}. Of
course, the notion of the ISCO is well defined in the test-particle
limit (\ie~for $M_2 \to 0$) and the approach of Hughes and Blandford
reproduces this limit by construction and provides reasonable results
for a large set of parameters. However, it also leads to incorrect
final spins (\ie~{\boldmath $S$}$_{\rm fin}/M^2_{\rm fin} > 1$) for
very rapidly rotating black holes and even at very small mass ratios.

Inspired in part by this work, Buonanno, Lehner, and
Kidder~\cite{Buonanno:07b} have more recently proposed a variant in
which the angular momentum of the final black hole is assumed to be
the sum of the individual spins and of the orbital angular momentum of
a test particle at the ISCO of a Kerr black hole with the \textit{same
  spin} parameter as that of the final black hole, \ie
\begin{equation}
\label{Sf_BKL}
M^2 \mathbf{a}_{\rm fin} = M_1 M_2
\mathbf{L}_{\rm tp}\,(|\mathbf{a}_{\rm fin}|,\theta_{\rm fin}) +
M_{1}^{2}\, \mathbf{a}_1 + M_{2}^{2}\, \mathbf{a}_2 \, .
\end{equation}
where {\boldmath $a$}$_{1,2}=${\boldmath $S$}$_{1,2}/M^2_{1,2}$ are
the two dimensionless spin vectors ($|${\boldmath $a$}$_{1,2}|\in
[0,1]$), $\theta_{\rm fin}$ is the angle of the final spin {\boldmath
  $a$}$_{\rm fin}$ with respect to the direction of the initial
orbital angular momentum of the binary, and {\boldmath $L$}$_{\rm
  tp}(|${\boldmath $a$}$_{\rm fin}|,\theta_{\rm fin})$ is a vector
whose direction is determined by assuming that the angle between
{\boldmath $L$}$_{\rm tp}$ and the total spin {\boldmath
  $S$}$_{1}+${\boldmath $S$}$_{2}$ maintains constant during the
inspiral, and whose magnitude is the dimensionless angular momentum at
the ISCO for a test particle moving around a black hole with spin
$|${\boldmath $a$}$_{\rm fin}|$ on an orbit with inclination
$\theta_{\rm fin}$ with respect to the equatorial plane.

Because {\boldmath $L$}$_{\rm tp}$ depends implicitly on the final
spin {\boldmath $a$}$_{\rm fin}$, equation~(\ref{Sf_BKL}) cannot be
solved analytically and needs to be computed numerically. Furthermore,
attention must be paid that {\boldmath $L$}$_{\rm tp}$ refers to a
prograde (retrograde) orbit if the final spin is aligned
(anti-aligned).

Perhaps the most surprising aspect of this approach is that despite
its simplicity, it is remarkably precise in a large portion of the
space of parameters, with differences from the numerical-relativity
results that can be as small as $\sim 1\%$ and of $\sim 10\%$ at
most. As I will comment later on, I regard this result as an evidence
that the dependence of the final spin on the initial conditions is
particularly simple and that, as a consequence, the mapping between
the initial and final state can be accomplished with rather simple
expressions. After all, {\boldmath $L$}$_{\rm tp}$ and {\boldmath
  $a$}$_{1,2}$ have a limited functional excursion and are they
linearly related in equation~(\ref{Sf_BKL}).

An important ingredient in the above recipe is ``mass conservation'',
\ie~the assumption that the mass of the final black hole is the same
as the sum of the two initial masses. Because gravitational radiation
carries away part of the mass of the system, such an assumption is
trivially incorrect, at least from a formal standpoint. However,
although the assumption is false, the resulting approximation is not
too severe. Numerical simulations reveal in fact that the mass
radiated in gravitational waves is $M_{\rm rad}/M = 1-M_{\rm fin}/M
\lesssim 5-7\times 10^{-2}$, with $M\equiv M_1+M_2$ being the total
mass. Although this difference is of the order of the precision with
which one wants to predict the final spin, a favourable series of
cancellations reduces the impact of this otherwise unreasonable
assumption.

In the spirit of improving the accord with the simulations and in
particular to produce more reliable predictions in the case of the
merger of maximally spinning black holes, Kesden~\cite{Kesden:2008}
has recently proposed an improvement over the approach in
ref.~\cite{Buonanno:07b} which can be used for binaries with spins
which are aligned or antialigned with the orbital angular momentum.
Once again using the test-particle motion in a Kerr spacetime, Kesden
has suggested that in addition to expression~(\ref{Sf_BKL}) being
valid, the final mass $M_{\rm fin}$ for arbitrary mass ratios and
initial spins can be computed as
\begin{equation}
\label{Mfin_K}
 M_{\rm fin} = M - \mu [1 - E(a_{\rm fin})] \, ,
\end{equation}
where $\mu \equiv M_1 M_2/M$ is the reduced mass and $E(a)$ is the
energy per unit mass of a test particle on an equatorial orbit of a
Kerr black hole with spin parameter $a$. This suggestion amounts to a
minimal change in the calculation of the final spin, but provides a
much more precise modelling of the final spin in the case of black
holes with very large spins, \ie~with $|${\boldmath
  $a$}$_1|\simeq|${\boldmath $a$}$_2| \gtrsim 0.99$.

The resulting picture is particularly simple for equal-spin black
holes, \ie~with $|${\boldmath $a$}$_1|=|${\boldmath $a$}$_2|=a$, (see
Fig. 2 of ref.~\cite{Kesden:2008}). In this case the test-particle
approach suggests that for $a \lesssim 0.9916$, the final spin $a_{\rm
  fin}$ is a function increasing monotonically with the symmetric mass
ratio $\nu\equiv M_1M_2/(M_1+M_2)^2=q/(1+q)^2$, so that equal-mass
mergers (\ie~with $\nu = 1/4$) are the most efficient ones in spinning
up the final black hole. However, for values of $a > 0.9916$ the
function $a_{\rm fin} (\nu)$ is no longer monotonically increasing
with $\nu$ but has a maximum at smaller and smaller values of $\nu$,
so that for these spins a suitable symmetric mass ratio $\nu \neq 1/4$
maximizes the final spin. This behaviour continues for increasingly
large values of $a$ and up to $a \sim 0.9987$, when $a_{\rm fin}
(\nu=1/4) = a$; as a result, for $a \gtrsim 0.9987$ equal-mass mergers
\textit{spin down} black holes rather than spinning them up,
\ie~$a_{\rm fin} (\nu=1/4) < a$ for $a \gtrsim 0.9987$. Finally, when
the initial black holes are maximally spinning, \ie~$a = 1$, the
maximum of $a_{\rm fin}(\nu)$ is at the extreme mass-ratio limit
(EMRL), \ie~at $\nu = 0$, thus making $a_{\rm fin}(\nu)$ monotonically
decreasing with $\nu$\footnote{Wald has shown in ref.~\cite{Wald:1974}
  that the cosmic censorship hypothesis holds also in the
  neighbourhood of the EMRL, \ie~that $a_{\rm fin} (\nu) \leq 1$ for
  $\nu\simeq 0$.}. As a direct consequence, maximally spinning black
holes cannot be produced by the merger of two black holes, the only
exception being offered by the EMRL. Rather, the merger of maximally
spinning equal-mass black holes yields a black hole with $a_{\rm fin}
\simeq 0.9988$\footnote{A different astrophysical route to the
  production of maximally spinning black holes is of course that of
  accretion and this was shown by Bardeen in
  ref.~\cite{Bardeen:1970}. The arguments made there apply also here
  if one is considering a binary coalescence in the neighborhood of
  the EMRL, \ie~for $\nu \simeq 0$.}.

An alternative and distinct line of research for the calculation of
the final spin has instead exploited numerical-relativity calculations
as soon as the first successful evolutions were
possible~\cite{Campanelli:2006uy, Campanelli:2006vp,
  Campanelli:2006gf}. While these first approaches essentially
represented fits to the numerical data (which were initially scarce)
they provided the first accurate description of the problem, albeit in
a small region of the space of parameters. Subsequent work focussed
instead on the derivation of analytic expressions which would model
the numerical-relativity data but also exploit as much information as
possible either from perturbative studies, or from the symmetries of
the system when this is in the weak-field
limit~\cite{Rezzolla-etal-2007, BoyleKesdenNissanke:07,
  BoyleKesden:07, Marronetti:2007wz, Rezzolla-etal-2007b,
  Rezzolla-etal-2007c}. In this sense, these approaches are not blind
fits of the data, but, rather, use the numerical-relativity data to
construct a physically consistent and mathematically accurate
modelling of the final spin.

The common ground shared by the approaches in this second class is
given by the \textit{assumption} that the final spin, when seen as the
function {\boldmath $a$}$_{\rm fin}=~${\boldmath $a$}$_{\rm
  fin}(${\boldmath $a$}$_{1}$, {\boldmath $a$}$_{2}, \nu)$, can be
expressed as a Taylor expansion around {\boldmath
  $a$}$_{1}=~${\boldmath $a$}$_{2}=\nu=0$. Given that
$\vert${\boldmath $a$}$_{1,2}\vert \leq 1$, this may seem as a
mathematically reasonable assumption and the expectation that the
series is convergent over the whole space of parameters as a
legitimate one. However, this remains an assumption and more work is
needed to make this mathematical assumption also a physically
reasonable one. Sharing this assumption, then different routes are
chosen to constrain the coefficients and these may invoke more
mathematically-based considerations, as those proposed in
ref.~\cite{BoyleKesdenNissanke:07, BoyleKesden:07, Tichy:2008}, or
more physically-based considerations, as those proposed in
refs.~\cite{Rezzolla-etal-2007, Rezzolla-etal-2007b,
  Rezzolla-etal-2007c}. Both approaches often reach the same
conclusions, but there is little doubt that the systematic approach
proposed in refs.~\cite{BoyleKesdenNissanke:07, BoyleKesden:07} offers
the advantage of being more easily expandable to capture higher-order
effects.

Here, however, I will concentrate on reviewing the approach which,
with a minimal number (4) of physically reasonable assumptions and
with a minimal number (5) of free coefficients to be fixed from the
numerical data, leads to a formula that can model generic initial spin
configurations and mass ratios, thus covering all of the 7-dimensional
space of parameters~\cite{Rezzolla-etal-2007, Rezzolla-etal-2007b,
  Rezzolla-etal-2007c}. As I will show in Sect.~\ref{sect2}, besides
being simple and physically motivated, it is also remarkably accurate
in reproducing the final spin of more than 150 simulations.

In essence, the approach developed in refs.~\cite{Rezzolla-etal-2007,
  Rezzolla-etal-2007b, Rezzolla-etal-2007c} (hereafter simply the
``AEI formula'') amounts to considering the dimensionless spin vector
of the final black hole as given by the sum of the two initial spins
and of a ``third'' vector parallel to the initial orbital angular
momentum when the binaries are widely separated. This ``third'' vector
is an intrinsic ``property'' of the binary (it will be shown below
that this is essentially the orbital angular momentum \textit{not}
radiated), thus depending on the initial spin vectors and on the black
holes mass ratio, but not on the initial separation. The formula for
the final spin then simply describes the properties of this vector (so
far only of its length) in terms of the initial parameters of the
binary and of a set of coefficients to be determined from a comparison
with numerical simulations.

Let us now consider in more detail how to derive such a formula. As
mentioned in the Introduction, four assumptions are needed in order to
make the problem tractable analytically and these are listed in what
follows.

\medskip
\textit{(i) The mass radiated to gravitational waves $M_{\rm rad}$ can
  be neglected} \textit{i.e.}, $M_{\rm fin} = M_1 + M_2$.
\smallskip

As mentioned before, while this assumption is certainly not correct,
its influence on the overall accuracy of the prediction is small, with
the possible exception of binaries with very small mass ratios and
with rapidly spinning black holes~\cite{Kesden:2008}. Work is in
progress to relax this assumption and will be presented in a
forthcoming paper~\cite{Barausse:2009}.

\medskip
\textit{(ii) At a sufficiently large but finite initial separation the
  final spin vector {\boldmath ${S}$}$_{\rm fin}$ can be approximated
  as the sum of the two initial spin vectors and of a third vector
  {\boldmath ${\tilde{\ell}}$} }
\begin{displaymath}
\label{assumption_1}
\mathbf{S}_{\rm fin}=\mathbf{S}_1+
\mathbf{S}_2+ \mathbf{\tilde{\ell}}\,,
\end{displaymath}
\smallskip

When viewed as expressing the conservation of the total angular
momentum, eq.~(\ref{assumption_1}) also defines the vector {\boldmath
  ${\tilde{\ell}}$} as the difference between the orbital angular
momentum when the binary is widely separated {\boldmath ${L}$}, and
the angular momentum radiated until the merger {\boldmath ${J}$}$_{\rm
  rad}$, \textit{i.e.}, {\boldmath ${\tilde \ell}$}$=${\boldmath
  ${L}$}$-${\boldmath ${J}$}$_{\rm rad}$. Stated differently, the
vector {\boldmath ${\tilde{\ell}}$} measures the orbital angular
momentum that \textit{cannot be radiated} and, assuming the
``effective'' ISCO as the radial separation at which the system
essentially stops radiating, {\boldmath ${\tilde{\ell}}$} can then be
thought of as the angular momentum of the binary at such ISCO.

\medskip
\textit{(iii) The vector {\boldmath ${\tilde{\ell}}$} is parallel to
  {\boldmath ${L}$}}.
\smallskip

This assumption (which is made also in ref.~\cite{Buonanno:07b})
essentially enforces that the component of the final spin {\boldmath
  ${S}$}$_{\rm fin}$ in the orbital plane equals the one of the total
initial spin {\boldmath ${S}$}$_1+${\boldmath ${S}$}$_2$ in that
plane.  This is correct when {\boldmath ${S}$}$_1=-${\boldmath
  ${S}$}$_2$ and $q=1$ [this can be seen from the PN equations at 2.5
  order], or by equatorial symmetry when the spins are aligned with
{\boldmath ${L}$}, or when {\boldmath ${S}$}$_1=${\boldmath
  ${S}$}$_2=0$ (also these cases can be seen from the PN
equations). However, for more general configurations one expects that
\boldell will also have a component orthogonal to {\boldmath ${L}$} as
a result, for instance, of spin-orbit or spin-spin couplings, which
will produce in general a precession of \boldell. In practice, the
component of \boldell orthogonal to {\boldmath ${L}$} will correspond
to the angular momentum {\boldmath ${J}$}$^{\perp}_{\rm rad}$ radiated
in a plane orthogonal to {\boldmath ${L}$}, with a resulting error in
the estimate of $\vert$ \boldell $\vert$ which is $\sim
\vert${\boldmath $J$}$^{\perp}_{\rm rad} \vert^2 / \vert${\boldmath
  ${\tilde{\ell}}$} $\vert^2\sim \vert$ {\boldmath
  ${J}$}$^{\perp}_{\rm rad} \vert^2/(2 \sqrt{3} M_1 M_2)^2$. A direct
consequence of assumption \textit{(iii)} is a systematic overestimate
of the final spin component along the direction of the orbital angular
momentum {\boldmath $a$}$^{\parallel}_{\rm fin}$ and, consequently, a
systematic overestimate of $\lesssim 2\%$ in the cosine of the
inclination angle $\theta_{\rm fin}$. In ref.~\cite{Tichy:2008} this
assumption was relaxed and the modelling of {\boldmath
  $a$}$^{\parallel}_{\rm fin}$ improved; however, a comparison with
the available data will show that the error introduced by this
assumption is not important for the estimate of the final spin
magnitude $\vert${\boldmath $a$}$\vert_{\rm fin}$ (see
Fig.~\ref{fig:residuals_all} and the discussion in Sect.~\ref{sect3}).

\medskip
\textit{(iv) When the initial spin vectors are equal and opposite
  ({\boldmath ${S}$}$_{1}=-${\boldmath ${S}$}$_{2}$) and the masses
  are equal ($q=1$), the spin of the final black hole is the same as
  for nonspinning binaries}.
\smallskip

Stated differently, equal-mass binaries with equal and opposite-spins
behave as nonspinning binaries, at least as far as the properties of
the final black hole are concerned. This condition is met by the
leading-order contributions to the spin-orbit and spin-spin
point-particle Hamiltonians and by the spin-induced radiation
flux~\cite{Barker1970,Buonanno:06cd}, but it is difficult to justify
in full general relativity.  Here it is introduced essentially to
break a possible degeneracy in the formula (see below) and it reflects
the expectation that if the spins are the same and opposite
(independently on whether they are aligned/antialigned with the
orbital angular momentum), their contributions to the final spin
cancel (at least to the precision of interest here) for equal-mass
binaries. Besides being physically reasonable, this expectation is met
by all of the simulations performed to date, both for spins aligned
with {\boldmath ${L}$}~\cite{Rezzolla-etal-2007,Rezzolla-etal-2007b}
and orthogonal to {\boldmath ${L}$}~\cite{Brugmann:2007zj,
  Marronetti:2007wz, Tichy:2008}. A similar assumption is also made,
although not explicitly, in ref.~\cite{Buonanno:07b}.

\bigskip

Using these assumptions it is now possible to derive the analytic
expression for the final spin. Let us start by expressing the vector
relation~(\ref{assumption_1}) as
\begin{equation}
\label{assumption_1_bis}
\mathbf{a}_{\rm
fin}=\frac{1}{(1+q)^2}
\left(\mathbf{a}_1+\mathbf{a}_2q^2 +
\mathbf{{\ell}} q \right)\,,
\end{equation}
where {\boldmath ${a}$}$_{\rm fin}=${\boldmath ${S}$}$_{\rm fin}/M^2$
[\textit{cf.} assumption \textit{(i)}], {\boldmath
  ${{\ell}}$}$\equiv$ {\boldmath ${\tilde{\ell}}$}$/(M_1 M_2)$,
{\boldmath ${a}$}$_{1,2}\equiv~$ {\boldmath ${S}$}$_{1,2}/M^2_{1,2}$,
and its norm is then given by

\begin{eqnarray}
\label{eq:general}
&\hskip -0.5cm \quad \vert \mathbf{a}_{\rm fin}\vert=
\frac{1}{(1+q)^2}\Big[ \vta{1}^2 + \vta{2}^2 q^4+
 2 {\vert \mathbf{a}_2\vert}{\vert
\mathbf{a}_1\vert} q^2 \cos \alpha +
\nonumber\\
& \hskip 2.5cm
2\left(
     {\vert \mathbf{a}_1\vert}\cos \beta +
     {\vert \mathbf{a}_2\vert} q^2  \cos \gamma
\right) {\vert \mathbf{{\ell}} \vert}{q}
+ \vert \mathbf{{\ell}}\vert^2 q^2
\Big]^{1/2}\,,
\end{eqnarray}
where the three vector cosine $\alpha, \beta$ and $\gamma$ are defined
by
\begin{equation}
\label{cosines}
\cos \alpha \equiv
{\mathbf{\hat{a}}_1\cdot\mathbf{\hat{a}}_2}
\,,
\hskip 0.3cm
\cos \beta \equiv
 \mathbf{\hat a}_1\cdot\mathbf{\hat{{\ell}}}\,,
\hskip 0.3cm
\cos \gamma \equiv
\mathbf{\hat{a}}_2\cdot\mathbf{\hat{{\ell}}}\,.
\end{equation}
Because {\boldmath ${a}$}$_{1,2}\parallel$ {\boldmath ${S}$}$_{1,2}$
and {\boldmath ${\ell}$} $\parallel$ {\boldmath ${L}$}, the angles
$\alpha, \beta$ and $\gamma$ are also those between the initial spin
vectors and the initial orbital angular momentum. Indeed, it is to
exploit this simplification that assumption \textit{(iii)} was
introduced and thus to replace {\boldmath ${\hat{a}}$}$_{1,2}$ with
{\boldmath ${\hat{S}}$}$_{1,2}$ and {\boldmath ${\hat{\ell}}$} with
{\boldmath ${\hat{L}}$} in~(\ref{cosines}). Note that in this line of
arguments I am making the underlying assumption that $\alpha, \beta$
and $\gamma$ are well-defined and hence that the initial separation of
the two black holes is sufficiently large that the the changes in
these angles due to precession is small. The error introduced in this
way in the measure of $\cos\alpha, \cos\beta$ and $\cos\gamma$ are
expected to be of the order of $\sim \vert${\boldmath
  $J$}$^{\perp}_{\rm rad} \vert / \vert${\boldmath ${\tilde{\ell}}$}$
\vert$.

All that is needed at this point is to ``measure'' $\vert${\boldmath
  ${\ell}$}$\vert$ by exploiting the results of numerical-relativity
simulations and, in particular, those for binaries with spins parallel
and aligned (\ie~$\alpha=\beta=\gamma=0$), for spins parallel and
antialigned spins (\ie~$\alpha=0$, $\beta=\gamma=\pi$), and for
antiparallel spins which are aligned or antialigned
($\alpha=\beta=\pi$, $\gamma=0$ or $\alpha=\gamma=\pi$, $\beta=0$).

As mentioned before, the matching with the numerical data is not
unique but this degeneracy can be broken if assumption \textit{(iv)}
is adopted. Doing this and requiring in addition that
$\vert${\boldmath $\ell$}$\vert$ depends linearly on $\cos\alpha$,
$\cos\beta$ and $\cos\gamma$, one finally obtains that the modulo of
the dimensionless, non-radiated angular momentum is given by (see
ref.~\cite{Rezzolla-etal-2007c} for a more detailed derivation of this
expression)
\begin{eqnarray}
\label{eq:L}
&&
\hskip -1.5cm
\vtl
 =
\frac{s_4}{(1+q^2)^2} \left(\vta{1}^2 + \vta{2}^2 q^4
	+ 2 \vta{1} \vta{2} q^2 \cos\alpha\right) +
\nonumber \\
&& \hskip 0.cm
\left(\frac{s_5 \nu + t_0 + 2}{1+q^2}\right)
	\left(\vta{1}\cos\beta + \vta{2} q^2 \cos\gamma\right) +
	2 \sqrt{3}+ t_2 \nu + t_3 \nu^2 \,.
\end{eqnarray}
The 5 coefficients $s_4, s_5, t_0, t_2, t_3$ appearing in (\ref{eq:L})
were determined by fitting a large number of binaries having
equal-mass and unequal but aligned/antialigned
spins~\cite{Rezzolla-etal-2007} or unequal-mass but equal
aligned/antialigned spins~\cite{Rezzolla-etal-2007b}. As a result of
this fitting process, the coefficients have been determined to be
\begin{eqnarray}
& s_4 \simeq -0.129 \,, \qquad \qquad \qquad & s_5 \simeq -0.384 \,, \nonumber\\
& t_0 \simeq -2.686 \,, \qquad \qquad \qquad & t_2 \simeq -3.454 \,, \nonumber\\
& t_3 \simeq~~~2.353 \,. &
\end{eqnarray}

Expressions~(\ref{eq:general}) and~(\ref{eq:L}) fully and accurately
determine the modulo of the final spin under generic spin
configurations and mass-ratio, but they do not yet provide information
about the different components of the final spin vector. Such
information can be extracted from the angle between the final spin
vector and the initial orbital angular momentum $\theta_{\rm fin}$,
which can be easily calculated from $|${\boldmath $a$}$_{\rm
  fin}|$. Because of assumption \textit{(iii)}, the component of the
final spin in the direction of {\boldmath ${L}$} is [\textit{cf.}
  eq.~(\ref{assumption_1_bis})]
\begin{equation}
\label{eq:a_z}
a_{\rm fin}^{\parallel}\equiv
\mathbf{a}_{\rm fin}\cdot \mathbf{\hat{{\ell}}}
= \frac{\vta{1} \cos\beta +
  \vta{2}q^2 \cos\gamma  + \vtl q}{(1+q)^2}\,,
\end{equation}
so that $\cos\theta_{\rm fin}={a_{\rm
    fin}^{\parallel}}/\vert${\boldmath $a$}$_{\rm fin}\vert$, and the
component orthogonal to the initial orbital angular momentum is
$a_{\rm fin}^{\perp}=$ {\boldmath $a$}$_{\rm fin}\sin\theta_{\rm fin}$. Note
that by construction our approach does not further decompose the
perpendicular component of the spin $a_{\rm fin}^{\perp}$.

Before moving on to a more detailed exploration in Sect.~\ref{sect3}
of the physical implications of expressions~(\ref{eq:general})
and~(\ref{eq:L}), it is useful to remark that the approach reviewed
here is intrinsically approximate. Although the comparison between the
predictions of the formula and all of the available numerical data has
revealed differences of a few percent at most, attention should be
paid to the fact that the coefficients have been constrained with a
rather small and statistically non-homogeneous set of
configurations. Most importantly, however, the formula discussed above
can be improved in a number of different ways which I list below:
\textit{(a)} by reducing the $\chi^2$ of the fitting coefficients as
new simulations are carried out; \textit{(b)} by using fitting
functions that are of higher-order than those used so far [\cf
  expressions~(\ref{eqmass_uneqspin}) and~(\ref{eqspin_uneqmass})] in
the spirit of the systematic spin expansion suggested in
ref.~\cite{BoyleKesdenNissanke:07, BoyleKesden:07}; \textit{(c)} by
estimating {\boldmath $J$}$^{\perp}_{\rm rad}$ through PN expressions
or by measuring it via numerical simulations; \textit{(d)} by
introducing an independent modelling of the mass radiated in
gravitational waves $M_{\rm rad}$. Work is in progress to include some
of these possible improvements and it will be presented
in~\cite{Barausse:2009}.

%-------------------------------------------------------------%
\section{Exploring the AEI spin formula}
\label{sect2}
%-------------------------------------------------------------%

In what follows I discuss in some detail the predictions of the AEI
spin formula for some simple cases and highlight how to extract
interesting physical considerations.

\subsection{Unequal mass, Aligned/Antialigned Equal spins}

If the black holes have \textit{unequal mass} but spins that are
\textit{equal, parallel} and \textit{aligned/antialigned} with the
orbital angular momentum, \ie~$|${\boldmath $a$}$_1|=|${\boldmath
  $a$}$_2|=a$, $\alpha=0\,; \beta=\gamma=0,\,\pi$, the prediction for
the final spin is given by the simple
expression~\cite{Rezzolla-etal-2007b}
\begin{equation}
\label{eqspin_uneqmass}
\fl \hskip 0.5cm
a_{\rm fin}(a,\nu)=a\cos\beta\left(1 + s_{4}a \nu + s_{5}\nu^2 + t_{0} \nu\right)+
2\sqrt{3}\nu+t_2\nu^2+t_{3}\nu^3\,.
\end{equation}
where $\cos\beta = \pm 1$ for aligned/antialigned spins. Note that
since the coefficients in (\ref{eqspin_uneqmass}) are determined by
fits to the numerical data and the latter is scarcely represented at
very large spins, the predictions of
expression~(\ref{eqspin_uneqmass}) for nearly maximal black holes are
essentially extrapolations and are therefore accurate to a few percent
at most. As an example, when $a=1$, the
formula~(\ref{eqspin_uneqmass}) is a non-monotonic function with
maximum $a_{\rm fin} \simeq 1.029$ for $\nu \simeq 0.093$; this
clearly is an artifact of the extrapolation.

\begin{figure}
\centerline{
\resizebox{11cm}{!}{\includegraphics[angle=-90]{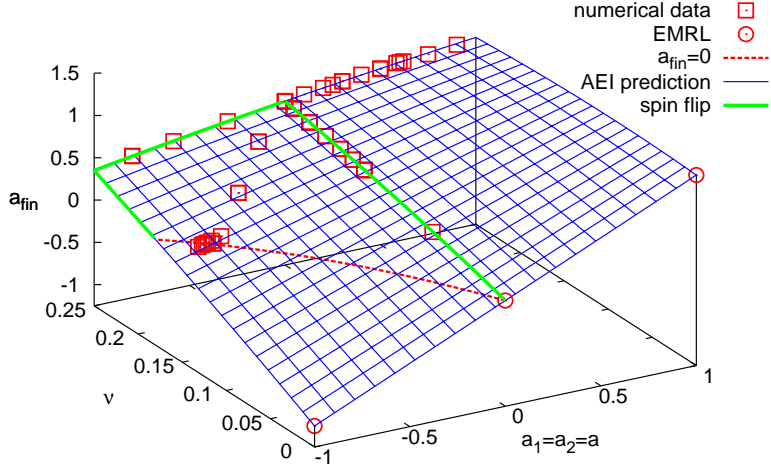}}
}
\vskip -0.5cm
\caption{Global dependence of the final spin on the symmetric mass
  ratio and on the initial spins as predicted by
  expression~(\ref{eqspin_uneqmass}) for equal-mass,
  aligned/antialigned equal-spin binaries. Squares refer to
  numerical-relativity values as reported in
  refs.~\cite{Campanelli:2006fy,Berti:2007snb, Buonanno:2007pf,
    Rezzolla-etal-2007b, Marronetti:2007wz,
    Berti:2007nw,Gonzalez:2008}, while circles to the EMRL
  constraints. Indicated with a (red) dashed line is the locus of
  points leading to a Schwarzschild black hole (\ie~$a_{\rm fin}=0$),
  while (green) solid lines mark the region leading to a ``spin-flip''
  (\ie~$a_{\rm fin}a < 0$).}
\label{fig:unequalmass_equalspin}
\end{figure}

The global behaviour of the final spin for unequal-mass and
aligned/antialigned equal-spin binaries is summarized in
Fig.~\ref{fig:unequalmass_equalspin}, which shows the functional
dependence of expression~(\ref{eqspin_uneqmass}) on the symmetric mass
ratio and on the initial spins. Squares refer to numerical-relativity
values as reported in refs.~\cite{Campanelli:2006fy,Berti:2007snb,
  Buonanno:2007pf, Rezzolla-etal-2007b, Marronetti:2007wz,
  Berti:2007nw,Gonzalez:2008}, while circles to the EMRL
constraints. A number of interesting considerations can now be made:

  \textit{(a)} Using~(\ref{eqspin_uneqmass}) it is possible to
  estimate that the minimum and maximum final spins for an equal-mass
  binary are $a_{\rm fin} = 0.3502 \pm 0.03$ and $a_{\rm fin} = 0.9590
  \pm 0.03$, respectively. While the value for the maximum spin is
  most likely underestimated (indeed Kesden predicts for this
  configuration $a_{\rm fin} = 0.9987$~\cite{Kesden:2008}) the minimum
  value is expected to be much more accurate than the estimate in
  ref.~\cite{Kesden:2008}, which tends to underestimate the final spin
  for $a \lesssim -0.3$ (\cf Fig. 5 of ref.~\cite{Kesden:2008}.

  \textit{(b)} Using~(\ref{eqspin_uneqmass}) it is straightforward to
  determine the conditions under which the merger will lead to a final
  \textit{Schwarzschild} black hole. In practice this amounts to
  requiring $a_{\rm fin}(a,\nu)=0$ and this curve is shown as a (red)
  dashed line in Fig.~\ref{fig:unequalmass_equalspin}. Several
  numerical simulations have been carried out to validate this
  condition~\cite{Rezzolla-etal-2007b,Berti:2007nw} and all of them
  have shown to produce black holes with $a_{\rm fin} \lesssim 0.01$
  (\cf squares in Fig.~\ref{fig:unequalmass_equalspin} with $\nu
  \simeq 0.16$). Overall, the behaviour captured by expression
  (\ref{eqspin_uneqmass}) shows that in order to produce a nonspinning
  black hole it is necessary to have unequal-masses (the largest
  possible mass ratio is $\nu \simeq 0.18$) and spins antialigned with
  the orbital angular momentum to cancel the contribution of the
  orbital angular momentum to the total one.

  \textit{(c)} Using~(\ref{eqspin_uneqmass}) it is also
  straightforward to determine the conditions under which the merger
  will lead to a \textit{``spin-flip''}, namely when the newly formed
  black hole will spin in the direction opposite to that of the two
  initial black holes. Mathematically this is equivalent to determine
  the region in the plane $(a, \nu)$ such that $a_{\rm fin}(a, \nu)\,a
  < 0$ and it is shown in Fig.~\ref{fig:unequalmass_equalspin} as
  limited by the (green) solid lines and by the (red) dashed
  line. Overall, it is clear that a spin-flip can take place only for
  very large mass ratios if the black holes are initially rapidly
  spinning and that small mass ratios will lead to a spin-flip only
  for binaries with very small spins.

  \textit{(d)} Finally, using~(\ref{eqspin_uneqmass}) it is also
  possible to determine the conditions under which the merger will
  lead to a final black hole with the \textit{same} spin as the
  initial ones. This amounts to requiring that $a_{\rm fin}(a,\nu)-a =
  0$ and only a very small portion of the $(a, \nu)$ plane does
  satisfy this condition (\cf Fig. 5 of
  ref.~\cite{Rezzolla-etal-2007b}). For equal-mass binaries, for
  instance, the critical value is $a_{\rm crit} \gtrsim 0.946$ and no
  spin-down is possible for $\nu \lesssim 0.192$. Because of the
  minuteness of the region for which $a_{\rm fin} < a$, black holes
  from aligned-spins binaries are typically \textit{spun-up} by
  mergers. As it will be shown also in the following Section, this
  statement is true also for other configurations and is probably true
  in general.

\subsection{Equal-mass, Aligned/Antialigned Unequal spins}

Equally interesting is to consider the prediction for the final spin
in the case in which the initial black holes have \textit{equal mass}
but \textit{unequal} spins that are either \textit{parallel} or
\textit{antiparallel} to the orbital angular momentum, \ie~for $q=1$
and $\alpha=0, \pi\,; \beta=0,\,\pi\,; \gamma=0,\,\pi$. Setting
$2|${\boldmath $a$}$_{1}|\cos\beta = a_1 + a_2$ in
expression~(\ref{eq:general}) we obtain the simple expression for the
final spin in these cases~\cite{Rezzolla-etal-2007}
\begin{equation}
\label{eqmass_uneqspin}
a_{\rm fin}(a_1,a_2)=p_0 + p_1 (a_1 + a_2) + p_2 (a_1 + a_2)^2\,,
\end{equation}
where the coefficients $p_0, p_1$ and $p_2$ are given by 
\begin{eqnarray}
\label{relations}
&p_0 = \frac{\sqrt{3}}{2} + \frac{t_2}{16} + \frac{t_3}{64}
	\simeq  0.6869\,, \qquad
&p_1 = \frac{1}{2} + \frac{s_5}{32} + \frac{t_0}{8}
        \simeq 0.1522\,, \nonumber\\
&p_2 = \frac{s_4}{16} \simeq -0.0081\,. &
\end{eqnarray}
Note that the coefficients $p_0, p_1, p_2$ and $s_4, s_5, t_0, t_2,
t_3$ were obtained through independent fits of two distinct data
sets. The fact they satisfy the conditions~(\ref{relations}) within
the expected error-bars is an important consistency check.

When seen as a power series of the initial spins,
expression~(\ref{eqmass_uneqspin}) suggests an interesting physical
interpretation. Its zeroth-order term, $p_0$, can be associated with
the (dimensionless) orbital angular momentum not radiated in
gravitational waves and thus amounting to $\sim 70\%$ of the final
spin at most. Interestingly, the value for $p_0$ is in very good
agreement with what is possibly the most accurate measurement of the
final spin from this configuration and that has been estimated to be
$a_{\rm fin}=0.68646 \pm 0.00004$~\cite{Scheel:2008}. Similarly, the
first-order term in (\ref{eqmass_uneqspin}), $p_1$, can be seen as the
contributions from the initial spins and from the spin-orbit coupling,
amounting to $\sim 30\%$ of the final spin at most. Finally, the
second-order term, $p_2$, can be seen as accounting for the spin-spin
coupling, with a contribution to the final spin which is of $\sim 4\%$
at most.

Another interesting consideration is possible for equal-mass binaries
having spins that are equal and antiparallel, \ie~$q=1$, {\boldmath
  $a$}$_1$ $=-$ {\boldmath $a$}$_2$. In this case,
expressions~(\ref{eq:general}) and~(\ref{eq:L}) reduce to
\begin{equation}
\label{2nd_check_bis}
\vtaf = \frac{\vtl}{4} = \frac{\sqrt{3}}{2} + \frac{t_2}{16} +
	\frac{t_3}{64}\,.
\end{equation}

Because for equal-mass black holes which are either nonspinning or
have equal and opposite spins, the vector {\boldmath $\ell$} does not
depend on the initial spins, expression~(\ref{2nd_check_bis}) states
that $\vert${\boldmath $\ell$}$\vert M_{\rm fin}^2/4=\vert${\boldmath
  $\ell$}$\vert M^2/4=\vert${\boldmath $\ell$}$\vert M_1 M_2$ is, for
such systems, the orbital angular momentum at the effective ISCO. We
can take this a step further and conjecture that $\vert${\boldmath
  $\ell$}$\vert M_1 M_2$ is the series expansion of the dimensionless
orbital angular momentum at the ISCO also for \textit{unequal-mass}
binaries which are either nonspinning or with equal and opposite
spins. The zeroth-order term of this series (namely, the term
$2\sqrt{3} M_1 M_2$) is exactly the one predicted from the EMRL.

\begin{figure}
\centerline{
\resizebox{8cm}{!}{\includegraphics[angle=-0]{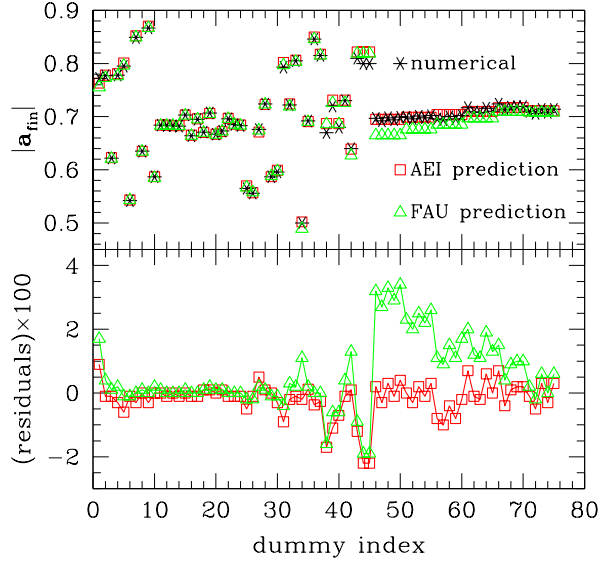}}
}
\vskip -0.5cm
\caption{\textit{Top panel}: Comparison between the final spin values
  $\vert${\boldmath $a$}$_{\rm fin}\vert$ obtained via numerical
  simulations and via the analytic formulas of
  AEI~\cite{Rezzolla-etal-2007c} (red squares) and of
  FAU~\cite{Tichy:2008} (green triangles). The numerical data has been
  published in ref.~\cite{Tichy:2008, Campanelli:2006vp,
    Campanelli:2007ew, Rezzolla-etal-2007c, Tichy:2007gso,
    Campanelli:2008} (see main text for a more detailed
  description).\textit{Bottom panel}: Residuals $\vert${\boldmath
    $a$}$_{\rm fin}\vert-\vert${\boldmath $a$}$_{\rm pred.}\vert$ for
  the AEI formulas (red squares) and for the FAU formula (green
  triangles).}
\label{fig:misaligned}
\end{figure}

\subsection{Generic (misaligned) binaries: unequal mass, unequal spins}

When the binaries are generic, namely when the initial spins are
oriented in generic directions and the two masses are different, the
spin formula~(\ref{eq:general})--(\ref{eq:L}) does not reduce to a
simple expression and the analysis of the physical implications
becomes more complex.

Much more challenging is also the numerical solution in these cases,
partly because they are computationally more expensive (no symmetries
can be exploited to reduce the computational domain), and partly
because the evolutions start at a finite separation which does not
account for the earlier evolution of the orbital angular momentum
vector and of the spins (both of which precess). In addition, because
the final spin is oriented in directions which are in principle
arbitrarily far from the main coordinate lines, the calculation of the
inclination angle from the properties of the final apparent horizon is
often non-trivial and suitable definitions need to be introduced (see,
\eg~\cite{Campanelli:2006fy}).

Notwithstanding this, accurate numerical simulations have been
reported by several groups and Fig.~\ref{fig:misaligned} reports data
for 75 of such binaries.  Using a ``dummy index'' to order the
different binaries, here is where the different results were
presented: the data from 1 to 31 was presented in
ref.~\cite{Tichy:2008}, the data from 32 to 34 in
ref.~\cite{Campanelli:2006vp,Campanelli:2007ew}, the data from 35 to
37 in ref.~\cite{Rezzolla-etal-2007c}, the data from 38 to 45 in
ref.~\cite{Tichy:2007gso}, and finally the data from 46 to 75 was
presented in ref.~\cite{Campanelli:2008}. The top panel of
Fig.~\ref{fig:misaligned}, in particular, reports a comparison between
the final spin values $\vert${\boldmath $a$}$_{\rm fin}\vert$ obtained
via numerical simulations and via the analytic formulas of
AEI~\cite{Rezzolla-etal-2007c} (red squares) and of
FAU~\cite{Tichy:2008} (green triangles). As mentioned before, the
latter has been proposed as a suitable application of the systematic
expansion suggested in~\cite{BoyleKesdenNissanke:07} when this is
truncated at first order. 

Because derived from a generic expansion and not from some assumptions
(as for the AEI formula), the the FAU formula offers the important
improvement of going beyond the restrictions imposed by our assumption
\textit{(iii)} and, in particular, models the changes of the final
spin orthogonal to the orbital angular momentum. I recall, in fact,
that as a consequence of assumption \textit{(iii)}, the final spin
component in the orbital plane, $a^{\perp}_{\rm fin}$ does not change
in the AEI formula and it is simply that of the initial total spin
{\boldmath $S$}$_{1}+${\boldmath $S$}$_{2}$. Tichy and
Marronetti~\cite{Tichy:2008}, instead, can model this change both via
a contraction (of about $70\%$) and a rotation (which is rather small,
with a rotation angle which is less than $9$ degrees). This leads to a
better estimate of the inclination angle\footnote{To fix the ideas and
  using as a reference the data published in
  ref.~\cite{Campanelli:2008}, the FAU formula has an average error on
  $\cos\theta_{\rm fin}$ of $\sim 0.1\%$, while the AEI formula has an
  average error of $\sim 1.7\%$.} but not necessarily of the final
spin modulo.

This is shown in the bottom panel of
Fig.~\ref{fig:misaligned}, where I report the residuals
$\vert${\boldmath $a$}$_{\rm fin}\vert-\vert${\boldmath $a$}$_{\rm
  pred.}\vert$ for the AEI formulas (red squares) and for the FAU
formula (green triangles). While the two approaches are overall
comparable, the FAU one has larger residuals in general. This may be
combined result of having used a small set of simulations to constrain
the coefficients (the FAU formula has in fact used only the data from
index 1 to index 31 and for these points the residuals are clearly
smaller) and of truncating the series expansion at first order (the
AEI formula has exploited a third-order expansion in the spins and
mass ratio).

\begin{figure}
\centerline{
  \resizebox{11cm}{!}{\includegraphics[angle=-90]{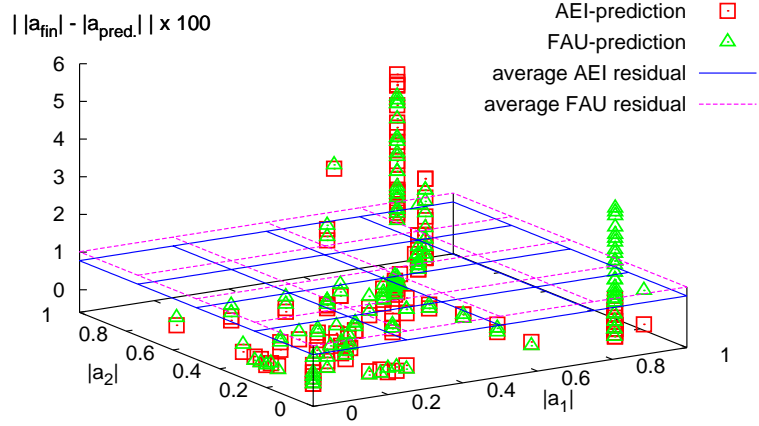}}
}
\vskip -0.5cm
\caption{Absolute residuals between the numerically computed value for
  the final spin and the one predicted by either the AEI formula (red
  squares) or the FAU formula (green triangles),
  \ie~$\vert\,\vert${\boldmath $a$}$_{\rm fin}\vert-\vert${\boldmath
    $a$}$_{\rm pred.}\vert\,\vert$. The numerical data is the one
  published in refs~\cite{Campanelli:2006fy,Berti:2007snb,
    Buonanno:2007pf, Rezzolla-etal-2007,Rezzolla-etal-2007b,
    Marronetti:2007wz, Berti:2007nw,Gonzalez:2008,Tichy:2008,
    Campanelli:2006vp, Campanelli:2007ew, Rezzolla-etal-2007c,
    Tichy:2007gso, Campanelli:2008,Scheel:2008,Gonzalez:2008} and is
  shown simply in terms of the initial spin moduli $|${\boldmath
    $a$}$|_{1,2}$.}
\label{fig:residuals_all}
\end{figure}

This conclusion applies also when comparing the two formulas against
all of the available numerical data, that is all of the data reported
for the 153 simulations presented in
refs.~\cite{Campanelli:2006fy,Berti:2007snb, Buonanno:2007pf,
  Rezzolla-etal-2007,Rezzolla-etal-2007b, Marronetti:2007wz,
  Berti:2007nw,Gonzalez:2008,Tichy:2008, Campanelli:2006vp,
  Campanelli:2007ew, Rezzolla-etal-2007c, Tichy:2007gso,
  Campanelli:2008,Scheel:2008,Gonzalez:2008}. This large bulk of data
is shown in a very synthetic form in Fig.~\ref{fig:residuals_all},
where I report the absolute residuals between the numerically computed
value for the final spin and the one predicted by either the AEI
formula (red squares) or the FAU formula (green triangles),
\ie~$\vert\,\vert${\boldmath $a$}$_{\rm fin}\vert-\vert${\boldmath
  $a$}$_{\rm pred.}\vert\,\vert$. Also indicated with constant planes
are the arithmetic average of the residuals for the AEI formula
(\ie~$0.775\times 10^{-2}$) and the arithmetic average of the
residuals for the FAU formula (\ie~$1.011\times 10^{-2}$).

A few remarks are worth making. Firstly, the two formulas yield very
similar results and while the AEI one is better on average, it also
has the largest residual and all for the data from
ref.~\cite{Herrmann:2007ex}. Secondly, configurations with very high
spins are those that produce the largest residuals and this is partly
due to the difficulty of having accurate calculations in those regimes
and, in part, to the fact that the coefficients in AEI's formula were
determined from simulations with moderate spins. Finally, and most
importantly, the data refers to simulations from different codes,
different numerical setups and hence different truncation errors. It
is therefore remarkable that the scatter is so small, with the largest
majority of residuals being below $1\%$. Stated differently,
Fig.~\ref{fig:residuals_all} can be interpreted as an absolute
variance of the precision of present numerical-relativity
calculations.

%-------------------------------------------------------------%
\section{Modelling the final recoil vector}
\label{sect3}
%-------------------------------------------------------------%

As mentioned in the Introduction, the final spin is not the only
quantity which is possible and useful to model from binary black-hole
coalescences. Rather, the gravitational recoil is a common feature of
binary black-hole coalescences and with multiple consequences. I
recall that a binary with unequal masses or unequal spins radiates
gravitational energy asymmetrically and the linear momentum lost at
any given time is not compensated by the loss after one period. As a
result, during the inspiral the linear momentum vector of the
asymmetric binary changes continuously with time increasing and
spiraling outwards up until the two black holes merge and the final
black hole conserves the linear momentum at the time of the
merger. This recoil velocity, or simply ``kick'', was predicted well
before numerical relativity simulations were able to prove its
existence and measure its magnitude~\cite{peres:1962,
  Bekenstein:1973mi}. Analytic estimates of the kick velocities have
also been available for some time~\cite{fitchett:1983,
  1984MNRAS.211..933F, Favata:2004wz, Blanchet:2005rj,
  Damour-Gopakumar-2006}, but because the largest part of the system's
acceleration is generated in the final parts of the last orbit, when
the motion is relativistic and the dynamics highly nonlinear, fully
relativistic calculations are necessary in order to determine the kick
accurately.

Indeed, modelling the recoil velocity has been one of the most
exciting results of recent numerical-relativity calculations, and
these have had a deep impact not only in general relativity but also
on astrophysics and cosmology. Discussing in detail such implications
would be very interesting but impossible with the space restrictions
of this contribution. Hereafter, therefore, I will only briefly review
the main results obtained over the last couple of years and provide a
snapshot of the present understanding of the recoil velocity from
binary black-hole coalescences.

Computations of the recoil velocity were made soon after the first
successful binary black hole evolutions and the systems which were
studied first were unequal-mass systems with moderate mass
ratios~\cite{Herrmann:2006ks,Baker:2006vn}. These initial studies were
quite limited in the mass ratios they were able to consider and were
further extended in ref.~\cite{Gonzalez:2006md}, where a large number
of models with mass ratios between $0.25$ and $1.0$ was explored. A
reasonably consistent picture has emerged from these simulations and
it is one in which the recoil velocities generated by mass asymmetries
are rather accurately bounded above by a maximum velocity of $\simeq
175\,\mathrm{km/s}$, which is attained for $\nu \simeq 0.195$, or
$q\simeq 0.36$. Although very small ratios (\ie~$q\lesssim 0.01$)
still represent the frontier of what is presently possible to compute
with standard finite-difference codes, the recent work of
ref.~\cite{Gonzalez:2008} with a mass ratio $q=0.1$ suggests that no
major surprise should be expected in the functional dependence of the
recoil velocity for $q < 0.1$.

As mentioned above, a recoil velocity is generated also when the
asymmetry is in the spins rather than in the masses. A number of
groups have therefore looked into the kick produced by equal-mass
binaries having, however, unequal spins~\cite{Herrmann:2007ex,
  Koppitz-etal-2007aa, Baker:2007gi, Campanelli:2007ew,
  Pollney:2007ss, Herrmann:2007ac}. In this context, particular
attention has been paid to binary black hole systems with equal-mass
and spins aligned with the orbital one, since systems of this type may
represent a preferred end-state of the binary evolution. PN studies in
vacuum have shown, in fact, that the gravitational spin-orbit coupling
has a tendency to align the spins when they are initially close to the
orbital one~\cite{Schnittman:2004vq}. Furthermore, if the binary
evolves in a disc, as expected for supermassive black holes, the
matter can exert a torque tending to align the
spins~\cite{Bogdanovic:2007hp}. Overall, the results of several groups
now agree that the recoil velocities generated from the merger of
equal-mass, aligned/antialigned unequal-spins are rather accurately
bounded above by a maximum velocity of $\simeq 450\,\mathrm{km/s}$,
which is attained for $\vert${\boldmath
  $a$}$_{1}\vert=-\vert${\boldmath $a$}$_{2}\vert=1$ (\cf~Fig. 2 of
ref.~\cite{Rezzolla-etal-2007}). Such velocities are large enough to
eject the merged black hole from the center or a dwarf elliptic or
spheroidal galaxy, and which have escape velocities $\lesssim
300\,\mathrm{km/s}$~\cite{Merritt:2004xa}.

Since the recoil is the result of an asymmetry in the binary system,
it does not come as a surprise that the largest recoils are those
produced in the most asymmetric systems, namely in those binaries
where the spins are antiparallel and lie in the orbital plane. These
are indeed the configurations leading to the ``superkicks'',
\ie~extremely large kicks which have been computed to be of the order
of $2000-2500\,\mathrm{km/s}$ and then extrapolated to yield
velocities as large as $4000\,\mathrm{km/s}$ for the maximally
spinning case~\cite{Campanelli:2007ew,
  Gonzalez:2007hi,Campanelli:2007cg,Brugmann:2007zj,Dain:2008}.  Such
velocities are so large to be able to eject the merged black hole from
the center of even massive elliptical galaxies (whose escape
velocities are $\lesssim 2000\,\mathrm{km/s}$~\cite{Merritt:2004xa})
and may represent a problem if this process does indeed take place
frequently during the merger of massive galaxies.

To clarify the picture of what contributes to what in the final
recoil, it is useful to adopt the phenomenological splitting first
proposed in ref.~\cite{Campanelli:2007ew}, and which distinguishes the
various contributions to the recoil velocity as coming from two
different sources of asymmetry in the system, namely the one relative
to \textit{mass-asymmetry} {\boldmath ${v}$}$_{_M}$ and the one
relative to the \textit{spin-asymmetry} {\boldmath ${v}$}$_{_S}$. The
latter is then further decomposed in the component contained in the
orbital plane (which is produced by the component of the spins
parallel to the orbital angular momentum) and in the component
parallel to the orbital angular momentum (which is produced by the
component of the spins orthogonal to the orbital angular momentum).
Selecting a Cartesian coordinate system in which the unit vector in
the $z$-direction {\boldmath ${e}$}$_{3}$ is aligned with the orbital
angular momentum {\boldmath ${L}$}, the above decomposition leads to a
generic expression
\begin{equation}
\label{v_kick}
{\bf v}_{\rm kick} = v_{_M} {\bf e}_1 +
v^{\perp}_{_S} (\cos\xi\, {\bf e}_1 + \sin\xi\, {\bf e}_2) +
v^{\parallel}_{_S} {\bf e}_3\,.
\end{equation}
It should be noted that expression (\ref{v_kick}) is essentially
heuristic although the different contributions can have approximate
expressions as derived from PN considerations. Over the last couple of
years a considerable effort has been invested in modelling the
different contributions which can now be expressed as follows. The
mass-asymmetry contribution is given
by~\cite{fitchett:1983,Gonzalez:2006md}
\begin{equation}
\label{v_kick_components_1}
v_{_M} = A \nu^2 (1 + B\nu)\sqrt{1 - 4\nu}\,, 
\end{equation}
the ``in-plane'', spin-asymmetry is given by~\cite{Herrmann:2007ex,
  Koppitz-etal-2007aa, Koppitz-etal-2007aa, Campanelli:2007ew,
  Pollney:2007ss, Herrmann:2007ac,Rezzolla-etal-2007}
\begin{equation}
\label{v_kick_components_2}
v^{\perp}_{_S} = \frac{\nu^2}{(1+q)}
 \left\vert C\left(a^{\parallel}_1 - q a^{\parallel}_2\right) +
      D\left[\left(a^{\parallel}_1\right)^2 -
          q^2\left(a^{\parallel}_2\right)^2\right]\right\vert\,, 
\end{equation}
while the ``off-plane'', spin-asymmetry is given
by~\cite{Gonzalez:2007hi, Campanelli:2007cg, Brugmann:2007zj,
  Lousto:2007db, Lousto:2008dn}
\begin{equation}
\label{v_kick_components_3}
v^{\parallel}_{_S} = E \frac{\nu^2}{(1+q)} \left\vert a^{\perp}_1 -
	q a^{\perp}_2\right\vert \cos(\Theta -\Theta_0) \,.
\end{equation}
Here, $A \simeq 1.2\times 10^4\,\mathrm{km/s}$, $B\simeq -0.93$, $C
\simeq 7040\,\mathrm{km/s}$, $D \simeq 1460\,\mathrm{km/s}$, and $E
\simeq 6.0\times 10^4\,\mathrm{km/s}$. Note that the angle $\Theta$ is
defined ``a-posteriori'' as the angle between the in-plane component
of the vector {\boldmath $\Delta$}$=(M_1+M_2)(${\boldmath $S$}$_1/M_1
-${\boldmath $S$}$_2/M_2)$ and the direction along which the two black
holes approach each other at the merger, while $\Theta_0$ is a simple
constant offset.

A number of considerations are in order. Firstly, it should be noted
that $v^{\perp}_{_S} \propto a^{\parallel}_i$ and viceversa
$v^{\parallel}_{_S} \propto a^{\perp}_i$, thus suggesting that the
unbalanced gravitational-wave emission (\ie~the one responsible for
the changes in of linear momentum vector) is always orthogonal to the
black holes spins. Secondly, the in-plane spin-asymmetry component
given by expression (\ref{v_kick_components_3}) shows that the largest
contributions come from binaries with anti-parallel initial spins,
\ie~$a^{\parallel}_1 = - a^{\parallel}_2$ and that, in this case, the
quadratic component is identically zero.  Thirdly, the mass dependence
of the off-plane spin-asymmetry component given by expression
(\ref{v_kick_components_3}) is presently under debate and indeed the
work carried out in ref.~\cite{Baker:2008md} suggests instead a
dependence of the type
\begin{equation}
\label{v_kick_components_3n}
v^{\parallel}_{_S} = E \frac{\nu^3}{(1+q)}
\left(
q a^{\perp}_2 \cos(\phi_2 - \Phi_2) - a^{\perp}_1 \cos(\phi_1 - \Phi_1)
\right) \,.
\end{equation}
where $\Phi_{1,2}$ are constant offsets and $\phi_{1,2}$ are the
angles between $a^{\perp}_{1,2}$ and some reference direction in the
equatorial plane. The most conspicuous and important difference
between expressions (\ref{v_kick_components_3}) and
(\ref{v_kick_components_3n}) is in the dependence on the symmetric
mass ratio (\ie~$\nu^2$ vs $\nu^3$) and this is not a minor difference
since it has important astrophysical consequences (see the discussion
in ref.~\cite{Baker:2008md}). It is presently unclear which of the two
expressions will find further confirmation and work is in
progress~\cite{Dorband:2009} to resolve this debate through an
independent set of simulations and a careful modelling of the
integration constant. The latter, I recall, accounts for the nonzero
initial linear momentum the spacetime has at the time the simulation
is started. An accurate determination of the integration constant has
played an important role in the determination of the second-order
correction in the contribution for $v^{\perp}_{_S}$ (see discussion in
ref.~\cite{Pollney:2007ss}).

%-------------------------------------------------------------%
\section{Conclusions}
\label{sect4}
%-------------------------------------------------------------%

The determination of the final spin and recoil velocity from the from
the knowledge of the initial properties of the initial black holes is
of great importance in several fields. In astrophysics, it provides
information on the properties of isolated stellar-mass black holes
produced at the end of the evolution of a binary system of massive
stars. In cosmology, it can be used to model the distribution of
masses and spins of the supermassive black holes produced through the
merger of galaxies~\cite{Berti2008,Arun:2008zn} and whether massive
galaxies should be expected to host a massive black hole at their
centre (but see also refs.~\cite{McNamara:2008dy,Sikora:2008fb}). In
addition, in gravitational-wave astronomy, the a-priori knowledge of
the final spin can help the detection of the
ringdown~\cite{Ajith:2007kx}.

I have reviewed those semi-analytic approaches to model the final
state from binary black-hole coalescences and that have been derived
to bypass the still very expensive numerical-relativity calculations
and obtain a complete and (reasonably) accurate description of the
full space of parameters. Although following different routes and
approximations, the several approaches proposed so far have provided a
rather accurate description of the final state from binary
coalescences. For the first time after decades, these approaches and
the numerical-relativity calculations behind them, have placed us in
the position of \textit{knowing} the properties of the final black
hole. However, our \textit{understanding} of these properties has not
progressed equally rapidly. It is now possible to compute the second
and third-order dependences on the initial spins and mass ratios but
many of these dependences are still unclear. Because knowledge without
understanding is often sterile, the works reported here should serve
as stimulus to other studies where the numerical results represent the
``background'' to be used by perturbative or PN studies to go from
knowledge to understanding. The recent works in
refs.~\cite{DamourNagar:07a, Berti:2007snb, Berti:2007nw, Baker:2008,
  Mino:2008, Favata:2008yd} are encouraging and useful steps in this
direction.

\ack
It is a pleasure to thank Peter Diener, Nils Dorband, Sascha Husa,
Denis Pollney, Christian Reisswig, Erik Schnetter and Jennifer Seiler
who have been my collaborators in this work. I am particularly
indebted to Enrico Barausse, who has been the source of many
discussions and ideas. This work was supported in part by the DFG
grant SFB/Transregio~7.

%-------------------------------------------------------------%
\section*{References}
%-------------------------------------------------------------%

%\begin{thebibliography}

\bibliographystyle{iopart-num}
\bibliography{aeireferences}

%\end{thebibliography}

\end{document}